\documentclass[prd,aps,eqsecnum,amsmath,amssymb,floatfix,nofootinbib,preprint,preprintnumbers,tightenlines]{revtex4}
\usepackage{amsmath,amssymb, bm}
\usepackage{dsfont}
\usepackage{epsfig}
\usepackage{graphicx}
\usepackage{slashed}
\usepackage{color}
\usepackage{subfigure}

\usepackage[colorlinks,citecolor=blue,urlcolor=blue,linkcolor=blue]{hyperref}

\newcommand{\be}{\begin{equation}}
\newcommand{\ee}{\end{equation}}

\newcommand{\beq}{\be}
\newcommand{\eeq}{\ee}
\newcommand{\bea}{\begin{eqnarray}}
\newcommand{\eea}{\end{eqnarray}}
\newcommand{\ba}{\begin{align}}
\newcommand{\ea}{\end{align}}
\newcommand{\bfig}{\begin{figure}}
\newcommand{\efig}{\end{figure}}

\newcommand{\wh}{\widehat}

\begin{document}

\font\rm=cmr12
\font\bf=cmbx12
\font\it=cmti12
\rm
\thispagestyle{empty}

\title{Revisiting the convergence of the perturbative QCD expansions based on conformal mapping of the Borel plane}

\vspace{3ex}
\author{Irinel Caprini}

\affiliation{Horia Hulubei National Institute for Physics and Nuclear Engineering, P.O.B. MG-6, 077125 Bucharest-Magurele,
 Romania\vspace{1cm}}

\begin{abstract}
\vskip0.5cm
 
The difference between fixed-order (FO) and contour-improved (CI) formulations of QCD perturbation theory limits the precision of the strong coupling determined from the hadronic decay of the $\tau$ lepton. Recently, several attempts to understand the mathematical origin of the difference and to solve it by subtracting the dominant infrared renormalon divergence have been made. Motivated by these studies, we review in this paper an improved perturbative QCD expansion, defined some time ago, which also exploits the renormalons by means of a suitable conformal mapping of the Borel plane. In particular,  we revisit the convergence of the new expansion, by completing the proof presented in a previous paper and showing that the domain of convergence is larger than stated before.  We also check the validity of the convergence conditions for the Adler function and  the CI and FO  expansions of the  $\tau$ hadronic spectral function  moments, and compare the approach based on conformal mapping  with recent solutions to the CIPT-FOPT discrepancy proposed in the literature.
\end{abstract}

\maketitle 
\vspace{0.2cm} 
%%%%%%%%%%%%%%%%%%%%%%%%%%%%%%%%%%%%%%%%%%%%%%%%%%%%%%%%%%%%%%%%%%%%%%%%%%%%%%%%%%%%%%%%%%%%%%%%%%
\section{Introduction}
%%%%%%%
 The hadronic decay of the  $\tau$ lepton is known to provide an important method of extracting the strong coupling $\alpha_s$ at a relatively low  scale, $m_\tau=1.777\,\text{GeV}$. However, there is a significant difference between results obtained using  the so-called  fixed-order (FOPT) and contour-improved (CIPT) QCD perturbation theory, such that analyses based on CIPT generally arrive at larger values of $\alpha_s(m_\tau^2)$ than those based on FOPT  \cite{PDG}. The inconsistency between these two  representations of the
QCD corrections limits the precision to which the strong coupling can be determined from this process.

A large number of works investigated this problem   during the last decades \cite{Braaten:1988hc,  Braaten:1991qm, LeDiberder:1992zhd, Davier:2008sk, Pich:2013lsa, Beneke:2008ad, Beneke:2012vb, Pich:2016bdg, Boito:2016oam, Hoang:2021nlz, Hoang:2020mkw, Benitez-Rathgeb:2022yqb, Benitez-Rathgeb:2022hfj, Gracia:2023qdy, Golterman:2023oml,  Beneke:2023wkq}.  An important point that cannot be overlooked in these  analyses is the fact, first pointed out by Dyson \cite{Dyson:1952tj}, that the perturbative expansions in quantum field theories  are divergent series, which can be at most asymptotic to the expanded functions. In QCD, the expansions of the Green functions are not only divergent, but also Borel non-summable \cite{tHooft}, because  some of the singularities in the Borel plane, the so-called infrared renormalons, prevent the unambiguous reconstruction of the original function by means of the Laplace-Borel integral \cite{Beneke:1998ui}. As a consequence, nonperturbative terms must be added to the perturbative series in order to obtain a definite result

In a series of recent papers \cite{Hoang:2021nlz, Hoang:2020mkw,  Benitez-Rathgeb:2022yqb, Benitez-Rathgeb:2022hfj,  Gracia:2023qdy, Golterman:2023oml, Beneke:2023wkq}, the difference between FO and CI expansions of the spectral moments in $\tau$ decay was shown to be due to a strong sensitivity to the infrared renormalons, especially the gluon condensate renormalon. Moreover, ways to resolve the discrepancy by subtracting the infrared renormalon divergence related to the gluon condensate have been proposed in \cite{Benitez-Rathgeb:2022yqb, Benitez-Rathgeb:2022hfj,  Beneke:2023wkq}. The goal is to reduce the main source of theoretical uncertainty and to improve the precision of $\alpha_s$ determination from hadronic $\tau$ decays.

In this context, it is useful to recall that modified perturbative expansions that also incorporate information about renormalons, and moreover have a tamed large-order behavior, can be obtained by the method of conformal mapping. It is known that by using a suitable conformal mapping one can accelerate the convergence of a power series and achieve its analytic continuation outside the original disk of convergence. In particle physics, the method was applied for the first time in \cite{CiFi, Frazer:1961zz} for the analytic continuation of hadronic scattering amplitudes. In QCD, it turns out that the method cannot be applied to the perturbative expansions of the Green functions in powers of the strong coupling, because they are singular at the expansion point. But the method can be used for the expansion of the Borel transform in the Borel plane. 

 The use of a conformal mapping of the Borel plane was suggested in \cite{Mueller:1992xz} and applied in  \cite{Altarelli:1994vz} as a technique to handle the ambiguities of the QCD perturbative series due to the large momenta in the Feynman integrals, which are harmless. The conformal mapping proposed in these works  takes into account only the ultraviolet renormalons. An important step forward was achieved in \cite{Caprini:1998wg}, where the optimal conformal mapping, which has the best convergence rate, was found for the QCD  the Adler function.  The optimal mapping transforms the whole Borel plane, with  cuts along the real axis due to both ultraviolet and infrared renormalons, into the unit disk in a new complex plane. In this framework, the Borel transform of the Adler function, or of the spectral moments, is expanded in powers of this optimal variable,  and  the expanded function is recovered from the Borel transform by Laplace-Borel integral regularized with the principal value (PV) prescription.

 The modified QCD perturbative expansions based on the conformal mapping of the Borel plane have been applied in \cite{Caprini:2000js, Caprini:2001mn, Cvetic:2001sn, Jeong:2002ph,  Caprini:2009vf, Caprini:2011ya, Abbas:2013usa, Caprini:2019kwp, Caprini:2020lff, Caprini:2021wvf}  to physical problems, in particular to the CI and FO expansions for the description of the  $\tau$-lepton decays. The mathematical properties of the modified expansions based on conformal mapping have been also investigated in detail.  Thus, in \cite{Caprini:2000js} it was shown that these expansions  converge when several conditions are fulfilled, in \cite{Caprini:2001mn} the properties of the expansion functions were investigated, and in  \cite{Caprini:2011ya}  the increase of the convergence rate by conformal mappings was 
demonstrated.
  
 In the present paper we revisit the proof of convergence given in  \cite{Caprini:2000js}. We complete and improve the arguments presented in  \cite{Caprini:2000js}, showing that the convergence domain is larger than previously stated. We also make some generalizations of interest for phenomenological applications. The work was motivated by the modified perturbative expansions based on renormalon subtraction, proposed recently in \cite{Benitez-Rathgeb:2022yqb, Benitez-Rathgeb:2022hfj,  Beneke:2023wkq}. We thought it may be of interest to bring into attention the modified expansions based on conformal mapping of the Borel plane, which exploit the renormalons in a different way. The  aim is to better understand the CIPT and FOPT expansions in this framework. 
 
The outline of the paper is as follows: in the next section we briefly review the FO and CI expansions of the Adler function and the moments of the $\tau$ hadronic spectral function. In Sec. \ref{sec:conf}, we define modified perturbative expansions based on the conformal mapping of the Borel plane. In Sec. \ref{sec:proof} we complete and generalize  the proof of convergence given in \cite{Caprini:2000js}, and in Sec. \ref{sec:applic} we check the validity of the convergence conditions for the CI and FO expansions of the Adler function and the moments.  Finally,  Sec. \ref{sec:conc} contains a summary of the work and a brief comparison  with recent related works. 

%%%%%%%%%%%%%%%%%%%%%%%%%%%%%%%%%%%%%%%%%%%%%%%%%%%%%%%%%%%%%%%%%%%%%%%%%%%%%%%%%%%%%%%%%%%%%%%%%%%%%%%%%%
\section{Adler function and spectral moments}\label{sec:Adler}
%%%%%%%%%%%%%%%%%%%%%%%%%%%%%%%%%%%%%%%%%%%%%%%%%%%%%%%%%%%%%%%%%%%%%%%%%%%%%%%%%%%%%%%%%%%%%%%%%%%%%%%%%%%%%%
 We  consider the reduced Adler function \cite{Beneke:2008ad}
\beq\label{eq:D}
\widehat{D}(s) \equiv 4 \pi^2 D(s) -1,
\eeq
where $D(s)=-s \,d\Pi(s)/ds$ is the logarithmic derivative of the invariant amplitude $\Pi(s)$ of the two-current correlation tensor.  From general principles of field theory, it is known that $\wh D(s)$ is an analytic function of real type, i.e. it satisfies the  Schwarz reflection property, $\wh D(s^*)=\wh D^*(s)$,  in the complex $s$ plane cut along the timelike axis for $s\ge 4 m_\pi^2$. 

 In QCD perturbation theory,  $\wh D(s)$ is expanded as  
\beq\label{eq:hatD}
\widehat{D}(s) =\sum\limits_{n\ge 1} [a_s(\mu^2)]^n \,
\sum\limits_{k=1}^{n} k\, c_{n,k}\, (\ln (-s/\mu^2))^{k-1},
\eeq
in powers of the renormalized strong coupling $a_s(\mu^2) \equiv \alpha_s(\mu^2)/\pi$, defined in a certain renormalization scheme (RS) at the renormalization scale $\mu$. 

The  coefficients  $c_{n,1}$ in (\ref{eq:hatD}) are obtained from the calculation of  Feynman diagrams, while  $c_{n,k}$ with $k>1$ are expressed in terms of  $c_{m,1}$ with $m< n$  and the perturbative coefficients $\beta_n$ of the $\beta$ function, which governs the variation of the QCD coupling with the scale $\mu$ in each RS:
 \be\label{eq:rge}
 -\mu\frac {d a_s}
{d\mu}\equiv \beta(a_s)=\sum_{n\ge 1}
\beta_n a_s^{n+1}. \ee

For large spacelike values $s<0$, one can choose in (\ref{eq:hatD}) the scale $\mu^2=-s$, and obtain the renormalization-group improved expansion
\beq\label{eq:hatD1}
\widehat{D}(s) =\sum\limits_{n\ge 1} c_{n,1}\, [a_s(-s)]^n,
\eeq
where $a_s(-s)\equiv \alpha_s(-s)/\pi$ is the running coupling. The expansions (\ref{eq:hatD})  and (\ref{eq:hatD1}) are often used also for complex values of $s$, outside the timelike axis $s>0$ where the QCD perturbation theory fails to describe the strong interactions of hadrons.  In these applications, in particular in the calculation of the spectral function moments, the perturbative expansions (\ref{eq:hatD}) and (\ref{eq:hatD1}) are traditionally called ``fixed-order perturbation theory'' (FOPT) and ``contour-improved perturbation theory'' (CIPT), respectively.

The Adler function has been calculated in the $\overline{\mbox{MS}}$ scheme to order $\alpha_s^4$  (see \cite{Baikov:2008jh} and references therein).  On the other hand, it is known that at high orders $n$, the coefficients increase factorially,  $c_{n,1}\sim  n !$  \cite{Beneke:1998ui}.  Therefore, the series (\ref{eq:hatD1})  has zero radius of convergence and can be interpreted only as an asymptotic expansion to $\widehat{D}(s)$ for $a_s\to 0$. This indicates the fact that the Adler function, viewed as a function of the strong coupling $a_s$, is singular at the origin $a_s=0$ of the coupling plane. 

In some cases, the expanded functions can be recovered  from their divergent expansions through Borel summation. The Borel transform of the Adler function is defined by the power series
\be\label{eq:B}
 B_{\widehat D}(u)= \sum_{n=0}^\infty  b_n\, u^n,
\ee
where the coefficients $b_n$ are related to the perturbative coefficients $c_{n,1}$ by 
\be\label{eq:bn}
 b_n= \frac{c_{n+1,1}}{\beta_0^n \,n!}..
\ee
Here we used the standard notation $\beta_0=\beta_1/2$, and in our convention $\beta_0=9/4$.

The large-order increase of the coefficients of the perturbation series is encoded  in the singularities of the Borel transform in the complex $u$ plane. In the present case, it is known that $B_{\wh D}(u)$ has singularities at integer values of $u$ for $u\ge 2$  and  $u\le-1$, denoted as infrared (IR) and ultraviolet  (UV) renormalons, respectively 
(we neglect the instantons, which are situated at larger $u>0$) \cite{Beneke:1998ui}.  In a specific limit of perturbative QCD, known as large-$\beta_0$ approximation \cite{Beneke:1992ch, Broadhurst:1992si, Beneke:1994qe},  the singularities are poles, but beyond this limit they are branch points. For our study it is important that some information is available on the  nature of the leading singularities: namely, near the first branch points $u=-1$ and $u=2$,  $B_{\widehat D}(u)$ behaves like
\be\label{eq:gammapowers}
 B_{\widehat D}(u) \sim \frac{r_1}{(1+u)^{\gamma_{1}}} \quad \text{and} \quad  B_{\widehat D}(u)  \sim \frac{r_2}{(1-u/2)^{\gamma_{2}}}, 
\ee
respectively, where the residues $r_1$ and $r_2$ are not known, but the exponents
$\gamma_1$ and $\gamma_2$ have been estimated from renormalization-group invariance \cite{Beneke:2008ad}.

From the definition (\ref{eq:B}), it follows that the function $\wh D(s)$ defined by (\ref{eq:hatD1}) can be recovered formally from the Borel transform  $B_{\widehat D}(u)$ by the Laplace-Borel integral  
\be\label{eq:Laplace}
\wh D(s)=\frac{1}{\beta_0} \,\int\limits_0^\infty  
\exp{\left(\frac{-u}{\beta_0 a_s(-s)}\right)} \,  B_{\wh D}(u)\, d u.
\ee
Actually, due to the singularities of $ B_{\wh D}(u)$ for $u\ge 2$, the  integral (\ref{eq:Laplace}) is not defined and requires a regularization. As shown in \cite{Caprini:1999ma}, the principal value (PV) prescription, where the integral (\ref{eq:Laplace}) is defined as the semisum of the integrals along two lines, slightly above and below the real positive axis $u\ge 0$, is consistent with some of the analytic properties of the true function $\wh D(s)$, in particular Schwarz reflection property and the absence of cuts on the spacelike axis $s<0$ ouside the Landau region. Therefore, we shall adopt this prescription in what follows. 
 
We shall consider also the spectral moments $M_i(s_0)$, defined as weighted integrals  of the spectral function $\mbox{Im} \Pi(s)$ along the finite range $0<s<s_0$ of the timelike axis. By exploiting the analytic properties of $\Pi(s)$, they can be expressed as integrals of the Adler function along a contour in the complex
$s$ plane, chosen for convenience to be the circle $|s|=s_0$:
\be\label{eq:Mi}
M_{i}(s_0)= \frac{1}{2\pi i} \!\!\oint\limits_{|s|=s_0}\!\! \frac{d s}{s} 
\omega_i(s/s_0) \widehat{D}(s),
\ee
where the weights $\omega_i(s)$ are analytic in the $s$ plane. In phenomenological applications to the hadronic $\tau$ decay, the usual choice is $s_0=m_\tau^2$, but lower values of $s_0$ have been also  considered.
  
  By inserting  in (\ref{eq:Mi}) the series  (\ref{eq:hatD}), one defines the FO  expansion 
\beq\label{eq:MiFO}
M_{i,\text{FO}}(s_0)= \sum\limits_{n\ge 1} [a_s(s_0)]^n \sum\limits_{k=1}^n \, k\, c_{n,k}\,J_{k-1, i},
\eeq
with
\be\label{eq:Jk}
J_{k,i}=\frac{1}{2\pi i} \!\!\oint\limits_{|s|=s_0}\!\! \frac{d s}{s} 
\omega_i(s/s_0) \ln^k(-s/s_0).
\ee

Alternatively, by inserting  in 
(\ref{eq:Mi}) the series  (\ref{eq:hatD1}), one defines the CI expansion 
\beq\label{eq:MiCI}
M_{i, \text{CI}}(s_0)= \sum\limits_{n\ge 1} c_{n,1}  \frac{1}{2\pi i} \!\!\oint\limits_{|s|=s_0}\!\! \frac{d s}{s} 
\omega_i(s/s_0) [a_s(-s)]^n,
\eeq
where the running coupling $a_s(-s)$ is computed by integrating along the circle the solution of the renormalization-group equation (\ref{eq:rge}), known at present to five loops \cite{Baikov:2016tgj}.

Borel representations for the moments can be derived also. By inserting  the Laplace-Borel representation (\ref{eq:Laplace}) of the Adler function into the integral (\ref{eq:Mi}) and permutting the integrals we obtain
\beq\label{eq:MiCIB}
M_{i, \text{CI}} = \frac{1}{\beta_0} \text{PV} \,\int\limits_0^\infty  d u\, B_{\wh D}(u)\, \frac{1}{2\pi}\int\limits_{0}^{2\pi} \,  d\phi\,\omega_i(e^{i\phi})\, e^{\frac{-u}{\beta_0 a_s(-s)}},
\eeq
where $-s=s_0 \exp(i(\phi-\pi))$. 

On the other hand, starting from the FO expansion (\ref{eq:MiFO}),  one can define the Borel transform
\be\label{eq:BMi}
 B_{M_{i}, \text{FO}}(u)= \sum_{n=0}^\infty  b_{n,i}\, u^n,
\ee
where 
\be\label{eq:bnMi}
 b_{n,i}= \frac{1}{\beta_0^n \,n!}\,\sum\limits_{k=1}^n \, k\, c_{n,k} \,J_{k-1, i}.
\ee
Then $M_{i, \text{FO}}$ is recovered from its Borel transform by the Laplace-Borel integral
\be\label{eq:MiFOB}
M_{i, \text{FO}}=\frac{1}{\beta_0} \text{PV} \,\int\limits_0^\infty  
\exp{\left(\frac{-u}{\beta_0 a_s(s_0)}\right)} \,  B_{M_i, \text{FO}}(u)\, d u,
\ee
where we adopted the PV prescription, anticipating the presence of singularities in the Borel transform  $B_{M_i, \text{FO}}(u)$  on the integration axis. 

 In the large-$\beta_0$ limit, the  Borel transform $B_{M_i, \text{FO}}(u)$ defined in (\ref{eq:BMi}) can be expressed in  a simple way in terms of the Borel transform   $B_{\wh D}(u)$ of the Adler function. The relation is found starting  from the CI representation  (\ref{eq:MiCIB}) and noting that the integral upon $\phi$ can  be performed exactly in the one-loop approximation of the coupling, when 
 \beq\label{eq:oneloop}
\frac{1}{\beta_0 a_s(-s)}=\frac{1}{\beta_0 a_s(s_0)} + \ln\left(\frac{-s}{s_0}\right), 
\eeq
the last term being equal to $i(\phi-\pi)$. Then, the comparison with (\ref{eq:MiFOB}) leads to 
\beq\label{eq:BMiBD}
B_{M_i, \text{FO}}(u) = \left[\frac{1}{2\pi} \int_0^{2\pi} d\phi \,\omega_i(e^{i\phi})\, e^{-iu(\phi-\pi)}\right] B_{\wh D}(u).
\eeq
 The integral can be calculated exactly for polynomial weights $\omega_i$, when (\ref{eq:BMiBD}) can be written as \cite{Beneke:2008ad, Caprini:2019kwp}
\beq\label{eq:BPolBD}
B_{M_i, \text{FO}}(u)= \frac{1}{\pi}\,\frac{\sin\pi u}{P_i(u)}\,B_{\wh D}(u),
\eeq
where $P_i(u)$ is a polynomial. For instance, for $\omega_i(s/s_0)=(s/s_0)^n$ one has $P_i(u)=(u-n)$, and for the weight  $\omega_{\tau}(s/s_0)=(1-s/s_0)^3 (1+s/s_0)$, which appears in the expression of $\tau$ hadronic width,  $P_i(u)=u(u-1)(u-3)(u-4)/12$ (for more examples see \cite{Caprini:2019kwp}).

From (\ref{eq:BPolBD}) it follows that $B_{M_i, \text{FO}}(u)$ inherits from $B_{\wh D}(u)$  the singularities at integer values of $u$. However, these singularities  are partly compensated by the zeros of $\sin\pi u$, except for those corresponding to the zeros of the polynomial $P_i(u)$. In particular, if this polynomial does not vanish at $u=-1$ and $u=2$ (as is the case with the kinematical weight $\omega_\tau$), the nature of the leading renormalons  of $B_{M_i, \text{FO}}(u)$, obtained from (\ref{eq:gammapowers}), is given by the exponents $\gamma_1-1$ and $\gamma_2-1$, respectively.

 Beyond the large-$\beta_0$ approximation, the exact
nature of the first singularities of $B_{M_i, \text{FO}}(u)$ cannot be
established exactly. Therefore, a conjecture is necessary in applications that exploit this nature. 
 
\vspace{-0.2cm}
 
%%%%%%%%%%%%%%%%%%%%%%%%%%%%%%%%%%%%%%%%%%%%%%%%%%%%%%%%%%%%%%%%%%%%%%%%%%%%%%%%%%%%%%%%%%%%%%%%%%%%%%%%%%%%%%%%%%%%
\section{Conformal mapping of the Borel plane}\label{sec:conf}

%%%%%%%%%%%%%%%%%%%%%%%%%%%%%%%%%%%%%%%%%%%%%%%%%%%%%%%%%%%%%%%%%%%%%%%%%%%%%%%%%%%%%%%%%%%%%%%%%%%%%%%%%%%%%%%%%%%%
The method of conformal mappings is known in mathematics as a technique for
“series acceleration”, i.e. for increasing the rate of convergence of power series. By
expanding a function in powers of the variable that maps its analyticity domain
onto a disk, the new series converges in a larger region, beyond the convergence
domain of the original expansion, and has an increased asymptotic convergence rate
compared to the original series inside this domain. 
The method can be applied actually only if the expanded function is analytic in a region around
the expansion point. Therefore, it cannot be used in QCD  for the standard
perturbative series in powers of the coupling, since the expanded functions are singular at
the origin of the coupling plane.  However,  the conditions of applicability are satisfied
by the Borel transforms, like the function $B_{\wh D}(u)$ defined in  (\ref{eq:B}).

\begin{figure}
\includegraphics[width=6.6cm]{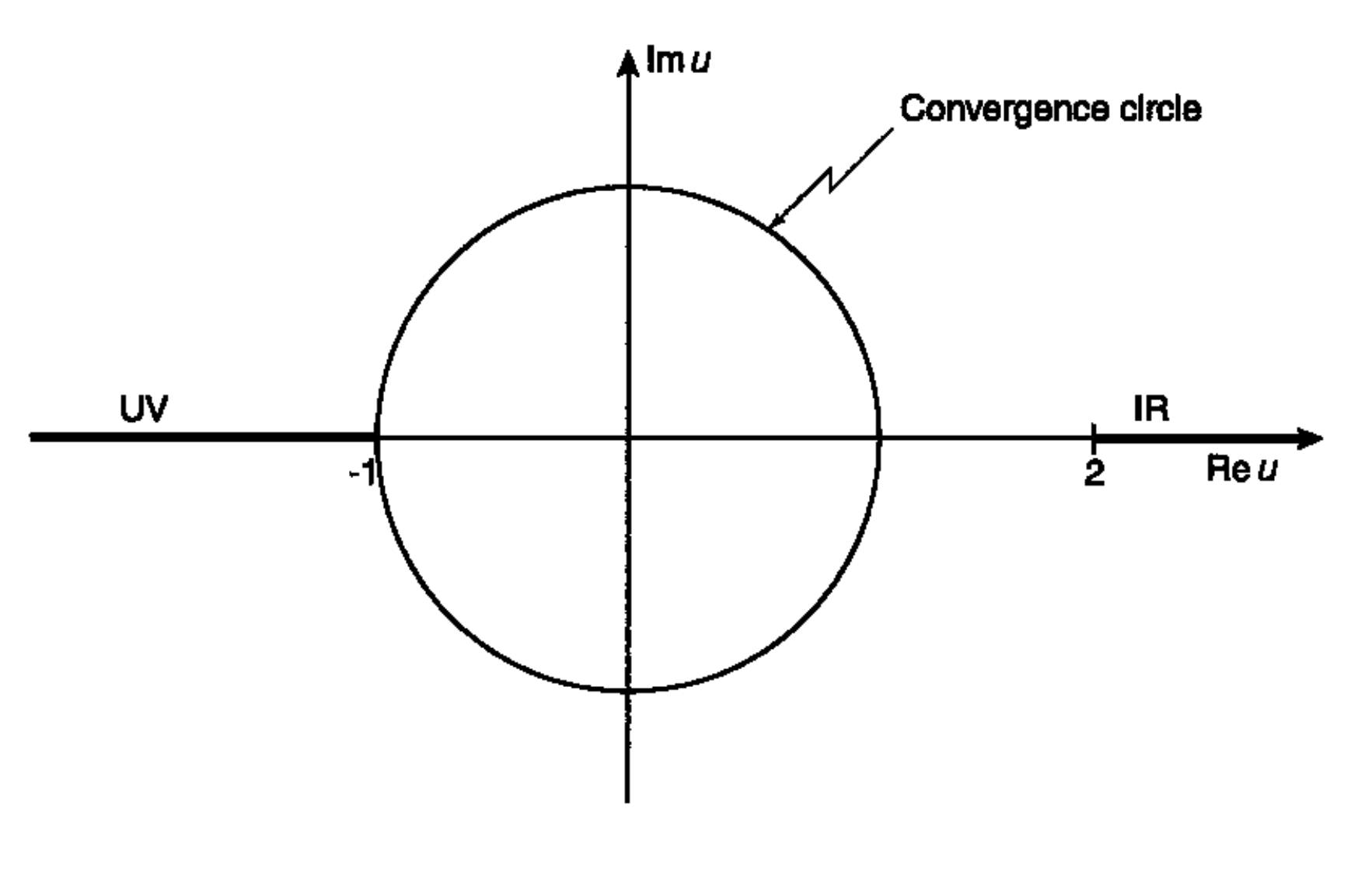}
\caption{Borel plane of the Adler function. The circle indicates the convergence domain of the series (\ref{eq:B}).  
\label{fig:Borel}}
\end{figure}

As indicated in Fig. \ref{fig:Borel}, the series (\ref{eq:B}) converges  in the disk $|u|<1$, limited by the first UV renormalon at $u=-1$. On the other hand, the Laplace-Borel integral (\ref{eq:Laplace}) includes the range  $u>1$, where the series (\ref{eq:B}) is divergent. This is the  reason of the divergence of the original series (\ref{eq:hatD1}), obtained formally by inserting  (\ref{eq:B}) in (\ref{eq:Laplace}) and integrating term by term.

%%%%%%%%%%%%%%%%%%%%%%%%%%%%%%%%%%%%%%%%%%%%%%%%%%%%%%%%%%%%%%%%%%%%%%%%%%%%%%%%%%%%%%%%%%%%%%%%%%%%%%%%%%%%%%%%%%%%%%%%%%%
%%%%%%%%%%%%%%%%%%%%%%%%%%%%%%%%%%%%%%%%%%%%%%%%%%%%%%%%%%%%%%%%%%%%%%%%%%%%%%%%%%%%%%%%%%%%%%%%%%%%%%%%%%%%%%%%%%%%%%%%%%%%%%%%%%%%%%%%%%%%

As discussed above, the domain of convergence  can be enlarged  by reexpanding  the function $B_{\wh D}(u)$ in powers of the variable which achieves the conformal mapping of the original complex $u$ plane onto the unit disk of a new complex plane. 
 This mapping, written for the first time in \cite{Caprini:1998wg},  has the form
\be\label{eq:w}
\tilde w(u)=\frac{\sqrt{1+u}-\sqrt{1-u/2}}{\sqrt{1+u}+\sqrt{1-u/2}},
\ee
and its inverse reads
\beq\label{eq:uw}
\tilde u(w)=\frac{8 w}{3-2 w+3 w^2}= \frac{8 w}{ 3 (w-\zeta) (w-\zeta^*)},
 \ee
where $\zeta= (\sqrt{2}+i)/(\sqrt{2}-i)$ and its complex conjugate  $\zeta^*$
are the images of $u=\infty$ on the unit circle in the $w$ plane.

\begin{figure}
\includegraphics[width=6cm, height=4.2cm]{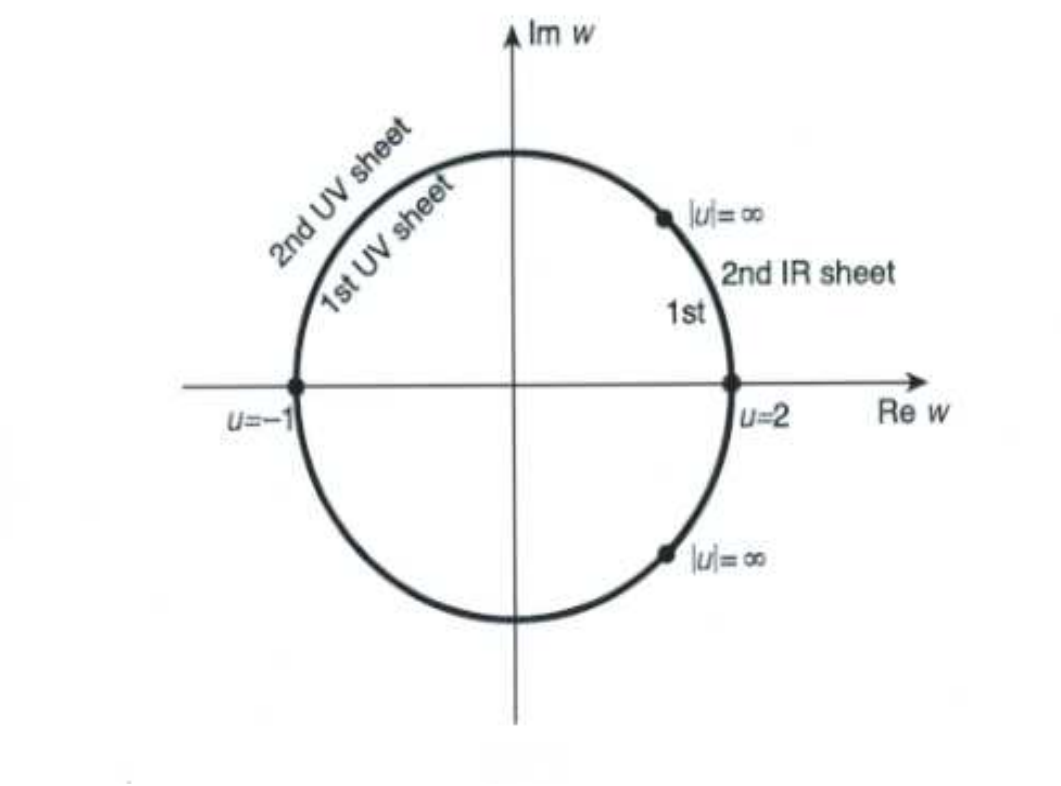}\hspace{0.5cm}
\caption{The  $w$ plane obtained by the conformal mapping (\ref{eq:w}). The IR and UV renormalons are mapped on the boundary of the unit disk.  
\label{fig:confBorel}}
\end{figure}

One can check that the function $\tilde w(u)$  maps the complex  $u$ plane cut along the real axis for $u\ge 2$ and $u\le -1$ onto the interior of the circle $\vert w\vert\, =\, 1$ in the complex plane $w\equiv \tilde w(u)$,  such that  the origin $u=0$ of the $u$ plane
corresponds to the origin $w=0$ of the $w$ plane, and the upper (lower) edges of the cuts are mapped onto the upper
(lower) semicircles in the  $w$ plane (see Fig. \ref{fig:confBorel}).  
By the  mapping (\ref{eq:w}), all  the singularities of the Borel transform, the  UV and IR  renormalons, are pushed to the boundary of the unit disk in the $w$  plane, at equal distance from the origin. 

Consider now the expansion of $B_{\widehat D}(u)$ in powers of the variable $w$:
\be\label{eq:Bw} B_{\widehat D}(u)=\sum_{n=0}^\infty c_n \, w^n, \quad\quad w \equiv \tilde w(u),\ee
 where the coefficients $c_{n}$ can be obtained from  the coefficients $b_{k}$,
$k\leq n$, using Eqs. (\ref{eq:B}) and  (\ref{eq:w}). By expanding $B_{\widehat D}(u)$ according to (\ref{eq:Bw}), one makes full use of its
holomorphy domain, because the known part of it
(the first Riemann sheet) is
mapped onto the convergence  disk.  Therefore, the series (\ref{eq:Bw}) converges in the whole $u$ complex plane up to the cuts, i.e., in a much larger domain than the original series (\ref{eq:B}).  Moreover, as shown in \cite{Caprini:2011ya},  this expansion has the best asymptotic convergence rate compared to other expansions, based on conformal mappings which map a part of the  holomorphy domain onto the unit disk.

%%%%%%%%%%%%%%%%%%%%%%%%%%%%%%%%%%%%%%%%%%%%%%%%%%%%%%%%%%%%%%%%%%%%%%%%%%%%%%%%%%%%%%%%%%%%%%%%%%%%%%%%%%%%%%%%%%%%%%%%%%%
%%%%%%%%%%%%%%%%%%%%%%%%
The expansion  (\ref{eq:Bw})  can be further improved by exploiting  the known behavior of the expanded function near the first branch points, discussed  below Eq. (\ref{eq:gammapowers}). This is done by expanding in powers of $w$ the product of  $B_{\wh D}(u)$ with a suitable factor $S(u)$, which compensates the singularities at $u=-1$ and $u=2$.  Actually, the product has still singularities (branch points)  at  $u=-1$ and $u=2$, generated by the terms of $B_{\wh D}(u)$ which are holomorphic at these points, but they are milder than the original ones (the singularities are ``softened'').  Therefore, the modified expansion
 \be\label{eq:Bw1}
 B_{\widehat D}(u)=\frac{1}{S(u)} \sum_{n=0}^\infty {\widetilde c}_n\, w^n
\ee
is expected to converge faster than the original expansion (\ref{eq:Bw}).

 As emphasized in  \cite{Caprini:2001mn, Caprini:2009vf, Caprini:2011ya}, while the optimal conformal mapping (\ref{eq:w}) is unique, the factorization of the singular factor $1/S(u)$ is not. We only require that $S(u)$ is analytic in the holomorphy domain of $B_{\widehat D}(u)$  and vanishes at $u=-1$ and $u=2$. Simple expressions, like 
 \be\label{eq:softu}  S(u)\sim (1+u)^{\gamma_1}(1-u/2)^{\gamma_2},\ee
 or
 \be\label{eq:softw}
  S(u)\sim (1+w)^{2\gamma_{1}} (1-w)^{2\gamma_{2}},\ee
have been investigated  in \cite{Caprini:2001mn, Caprini:2009vf, Caprini:2011ya}.
 
 In a similar way, we consider expansions in powers of  $w$ of the Borel transform $B_{M_i, \text{FO}}$ of the FO expansion of the moments, defined  in (\ref{eq:BMi}). 
 In the large-$\beta_0$ approximation, using (\ref{eq:BPolBD}) and the expansion (\ref{eq:Bw1}), we have
 \be\label{eq:BMiw}
 B_{M_i, \text{FO}}(u)=\frac{\sin\pi u}{\pi P_i(u) S(u)} \sum_{n=0}^\infty {\widetilde c}_n\, w^n,
 \ee
 where $P_i(u)$ is a polynomial, $S(u)$ the softening factor and  ${\widetilde c}_n$  the coefficients of the expansion  (\ref{eq:Bw1}).
 
 In the general case, beyond the one-loop approximation, by expanding in powers of $w$  the Borel transform defined in (\ref{eq:BMi}), we write
 \be\label{eq:Miw}
 B_{M_i, \text{FO}}(u)=\sum_{n=0}^\infty \hat c_{n,i}\, w^n, 
 \ee
 where $\hat c_{n,i}$ are obtained from the coefficients $b_{n,i}$ defined in  (\ref{eq:bnMi}). One can include also a softening factor, as for the Adler function in (\ref{eq:Bw1}), with a suitable assumption about the nature of the first singularities, as mentioned at the end of Sec. \ref{sec:Adler}.  

By inserting the  expansions (\ref{eq:Bw}) or (\ref{eq:Bw1})  of the Borel transform  in the Borel-Laplace integral (\ref{eq:Laplace}),  we obtain new perturbative series for the Adler function in the complex $s$ plane. For convenience, we use below the notation from \cite{Caprini:2000js},  writing
\be\label{eq:DI} 
\wh D(s)= \frac{1}{\beta_0} I(\beta_0 a_s(-s)).
\ee
Here $I$ denotes the series
\be\label{eq:cI}
I(a)=\sum_{n=0}^\infty c_n I_n (a),
\ee
where the coefficients $c_n$ appear in (\ref{eq:Bw}) and the expansion functions are
\be\label{eq:In}
I_n(a)=\text{PV} \int_0^\infty w^n e^{-u/a} du.
\ee
Alternatively, 
\be\label{eq:cItilde}
I(a)=\sum_{n=0}^\infty \widetilde c_n \widetilde I_n (a),
\ee
where the coefficients $\widetilde c_n$ appear in (\ref{eq:Bw1}) and the expansion functions are
\be\label{eq:Intilde}
\widetilde I_n(a)=\text{PV} \int_0^\infty \frac{w^n}{S(u)} e^{-u/a} du.
\ee

Returning to moments, the CI version is obtained 
by inserting the above expansions of the Adler function in the contour integral (\ref{eq:Mi}). For instance, using (\ref{eq:DI})-(\ref{eq:In}), we write
\be\label{eq:MicI}
M_{i, \text{CI}}(s_0)= \sum\limits_{n\ge 1} c_{n}  \frac{1}{2\pi i} \!\!\oint\limits_{|s|=s_0}\!\! \frac{d s}{s} 
\omega_i(s/s_0) I_n(\beta_0 a_s(-s)),
\ee
where we permuted the order of integration and summation since, as we will show below,  the series (\ref{eq:cI}) is absolutely convergent.

 In the FO version, the moments are expressed as 
\be\label{eq:MiI}
M_{i, \text{FO}}(s_0)=\frac{1}{\beta_0}\,I(\beta_0 a_s(s_0)),
\ee
where, in the large-$\beta_0$ approximation obtained from (\ref{eq:BMiw}), $I$ denotes the series
\be\label{eq:cIhat}
I(a)=\sum_{n=0}^\infty \widetilde c_n \widehat I_n (a),
\ee
with coefficients $\widetilde c_n$ defined in (\ref{eq:Bw1}) and expansion functions 
\be\label{eq:Inhat}
\widehat I_n(a)=\text{PV} \int_0^\infty \frac{\sin\pi u}{\pi P_i(u) S(u)} \, w^n\,e^{-u/a} du,
\ee
while in the general case
\be\label{eq:cIhati}
I(a)=\sum_{n=0}^\infty \hat c_{n,i} I_n (a),
\ee
where the coefficients  $\hat c_{n,i}$ appear in the expansion (\ref{eq:Miw}) and the expansion functions $I_n$ are defined in (\ref{eq:In}).

The analytic properties of the expansion functions defined above have been discussed in detail in \cite{Caprini:2001mn}, where it was shown that the functions $I_n(a)$ (denoted there as $W_n(a)$), are analytic in the complex $a$ plane and bounded for $\text{Re}\, a>0$, but exhibit a cut along the axis $a<0$ and an essential singularity  ($\sim \exp(-1/a))$ at the origin $a=0$. As a consequence, when expanded in powers of $a$, $I_n(a)$ have divergent expansions, with coefficients exhibiting factorial growth. On the other hand, as we will show in the next section, the expansion (\ref{eq:cI}) is convergent under certain conditions.

%%%%%%%%%%%%%%%%%%%%%%%%%%%%%%%%%%%%%%%%%%%%%%%%%%%%%%%%%%%%%%%%%%%%%%%%%%%%%%%%%%%%%%%%%%%%%%%%%%%%%%%%%%%%
\section{Convergence of the modified expansions}\label{sec:proof}
%%%%%%%%%%%%%%%%%%%%%%%%%%%%%%%%%%%%%%%%%%%%%%%%%%%%%%%%%%%%%%%%%%%%%%%%%%%%%%%%%%%%%%%%%%%%%%%%%%%%%%%%%%%%
\subsection{Method of steepest descent}\label{sec:steep} 
%%%%%%%%%%%%%%%%%%%%%%%%%%%%%%%%%%%%%%%%%%%%%%%%%%%%%%%%%%%%%%%%%%%%%%%%%%%%%%%%%%%%%%%%%%%%%%%%%%%%%%%
We first briefly review the main steps of the method of steepest descent applied in \cite{Caprini:2000js} for the estimation of the quantities $I_n(a)$ at large $n$. We recall that the Borel transform $B_{\wh D}(u)$ is a function  
of  real type, which satisfies $B_{\wh D}^*(u)=B_{\wh D}(u^*)$. Therefore, the coefficients $b_n$ of the expansion  
(\ref{eq:B}), as well as the coefficients $c_n$ of the expansion  
(\ref{eq:Bw}) are real. We consider the expansion (\ref{eq:cI})
for complex values of $a$ of the general form $a=|a| e^{i\psi}$, where $\psi = 
\arg a$ is the phase of $a$.

By writing the PV prescription in an explicit way, we first express (\ref{eq:In}) as
 \be\label{eq:Inpm}
I_{n}(a)=\frac{1}{2}\int\limits_{\cal C_+}\, {\rm e}^{-\frac{u}{a}}\, (\tilde w(u))^n
\, du +\frac{1}{2} \int\limits_{\cal C_-}\, e^{-\frac{u}{a}}\, (\tilde w(u))^n
\,du,
\ee
for $n=0,1,2,...$, where ${\cal C_+}$ (${\cal C_-}$) are lines parallel to the 
real positive axis, slightly above (below) it, and $\tilde w(u)$ is defined in (\ref{eq:w}).

The contribution to (\ref{eq:Inpm}) of the
integral along the contour ${\cal C}_+$ can be written as
 \be\label{eq:Inp}
I^+_n(a)=\int\limits_{{\cal C}_+}  e^{-F_n(u)}\, d u,
\ee
  where
\be\label{eq:Fn}
 F_n(u)=\frac{u}{a}- n \ln \tilde w(u).\ee
We evaluate the integral (\ref{eq:Inp}) for large $n$ by applying the method of 
steepest descent \cite{Jeff}. The saddle points are given by  the 
equation
\be\label{eq:deriv}
\frac{\tilde w'(u)}{\tilde w(u)}=\frac{1}{a n},
\ee 
which has four solutions, having at large $n$ the form
\be\label{eq:saddle}
 \frac{1+i}{ 2^{1/4}}  \sqrt{a n}\,,\,\,\, \frac{ 1-i}{ 2^{1/4}}  \sqrt{a n}\,,\,\,\,
\frac{-1+i}{ 2^{1/4}} \sqrt{a n}\,,\,\,\,\frac{-1-i}{ 2^{1/4}} \sqrt{a n}.
\ee 
Of interest for the evaluation of (\ref{eq:Inp}) is the point $u_0$ closest to the line ${\cal C_+}$
\be\label{eq:u0p}
u_0= 2^{-1/4} ( 1+i) \sqrt{a n}=|u_0| e^{i\alpha},\ee
with
\be\label{eq:u0p1}
|u_0|= 2^{1/4} \sqrt{|a| n},\quad\quad \alpha=\frac{\pi}{4}+\frac{\psi}{2}.
\ee
In our application to the Adler function, $a=\beta_0 a_s(-s)$, and from (\ref{eq:oneloop}) it follows that in the one-loop limit  $\cos \psi>0$, which means that
\be\label{eq:cond}
\vert\psi\vert <\frac{\pi}{2}.
\ee
Therefore, the point $u_0$ defined in (\ref{eq:u0p}) and (\ref{eq:u0p1})  is situated in the first quadrant of the $u$ plane. 
 
In order to evaluate the integral (\ref{eq:Inp}), we first rotate the contour ${\cal C}_+$ in 
the trigonometric direction in the upper half-plane, until it becomes a line ${\cal C}_+'$ 
passing through the origin and the saddle point $u_0$.
The rotation is possible since the function $\tilde w(u)$ has no 
singularities outside the real axis, and the arc of the circle at infinity 
gives a vanishing contribution, as can be easily verified.  Near the point $u_0$, $F_n(u)$ can be expanded as
\be\label{eq:steep}
F_n(u)=F_n(u_0)+\frac{1}{2}F''_n(u_0) (u-u_0)^2+....
\ee 
By using the expansion of $\tilde w(u)$ for large $u$ in the upper half plane 
($\tilde w(u)\approx \zeta (1-i\sqrt{2}/u)$, where $\zeta =(\sqrt{2}+i)/(\sqrt{2}-i)$), 
 we obtain after a straightfoward calculation
\bea\label{eq:Fn0}
e^{-F_n(u_0)}&&\approx\zeta^n\, \left(1-\frac{2^{3/4}i}{ (1+i)\sqrt{a n}}
\right)^n e^{-2^{-1/4}(1+i) \sqrt{\frac{n}{a}}}\nonumber \\
&&\approx  \zeta^n\
e^{-2^{3/4}(1+i) \sqrt{\frac{n}{a}}},
\eea 
and 
\be\label{Feq:2n0}
F_n''(u_0)\approx \frac{2^{1/4}(1-i)}{\sqrt{n a^3}}\,= |F_n''(u_0)|e^{i \beta},
\ee
where
\be\label{eq:beta} |F_n''(u_0)|=\frac{2^{3/4}}{\sqrt{n |a|^3}},\quad\quad
\beta=-\frac{\pi}{4}-\frac{3\psi}{2}.\ee
Then (\ref{eq:Inp}) becomes
\be\label{eq:Inp1}
I^+_n(a)\approx \zeta ^n  e^{-2^{3/4}(1+i) \sqrt{\frac{n}{a}}}\int\limits_
{{\cal C}_+'}
 e^{-\frac{ |F_n''(u_0)|}{ 2}\,e^{i\beta}\,(u-u_0)^2}\,d u.
\ee 
We further deform  the integration line into the path of steepest descent
without going outside the two valleys near the saddle point  $u_0$, by taking
\be\label{eq:steepline}
u- u_0 \approx \sqrt{2 /|F_n''(u_0)|} \,e^{-i\beta/2}\,\rho\,
\ee
with real $\rho$. The phase of $(u-u_0)^2$ exactly compensates the phase
of $F''_n(u_0)$, making
 the exponent of the integrand in (\ref{eq:Inp1}) real. The integrand  can be 
 written as $e^{-\rho^2}$ and the integral done explicitly gives
\be\label{eq:Inp2}
I^+_n(a)\approx \zeta ^n  e^{-2^{3/4}(1+i) \sqrt{\frac{n}{a}}}
 \frac{e^{-i\beta/2}}{
 \sqrt{ |F_n''(u_0)|/ 2}}  \,\frac{\sqrt{\pi}}{ 2},
\ee 
i.e. up to a constant independent of $n$
\be\label{eq:Inpsaddle}
I^+_n(a)\approx n^{\frac{1}{4}} \zeta^n
 e^{-2^{3/4} (1+i) \sqrt{\frac{n}{a}}}.
\ee

The evaluation of the integral along the contour ${\cal C}_-$ in (\ref{eq:Inpm})
proceeds in a  similar way. The saddle point of interest is 
\be\label{eq:u0m}
u'_0=2^{-1/4} (1- i) \sqrt{an}\,= \, 2^{1/4} \sqrt{|a| n}\,
e^{-(i\frac{\pi}{4}-\frac{\psi}{2})},
\ee
which is situated in the fourth quadrant of the complex $u$ plane for $\psi$ in the range
given in (\ref{eq:cond}). Instead of (\ref{eq:Inp1}), we have now
\be\label{eq:Inm1}
I^-_n(a)\approx (\zeta^*) ^n  e^{-2^{3/4}(1-i) \sqrt{\frac{n}{a}}}
\int\limits_{{\cal C}_-'}
 e^{-\frac{ |F_n''(u'_0)|}{2}\,e^{ i\beta'}\,(u-u'_0)^2}\,d u\,,
\ee 
where $\beta'=\pi/ 4-3\psi/ 2$ and ${\cal C}_-'$ is a contour rotated in the lower half-plane up to the point $u_0'$, which we further
 deform into the steepest descent path to obtain 
\be\label{eq:Inmsaddle}
I^-_n(a)\approx  n^{\frac{1}{4}} (\zeta^*)^n e^{-2^{3/4} (1- i) \sqrt{\frac{n}{a}}}.
\ee
Adding the two terms written  in (\ref{eq:Inpsaddle}) and (\ref{eq:Inmsaddle}), we obtain the large-$n$ behavior   
\be\label{eq:Insaddle}
I_n(a)\approx  n^{\frac{1}{4}} \zeta^n
 e^{-2^{3/4} (1+i) \sqrt{\frac{n}{a}}} +  n^{\frac{1}{4}} (\zeta^*)^n e^{-2^{3/4} (1- i) \sqrt{\frac{n}{a}}}
\ee
of the functions defined in (\ref{eq:Inpm}). In the next subsection we shall use the above estimate  for discussing the convergence of the series (\ref{eq:cI}).

%%%%%%%%%%%%%%%%%%%%%%%%%%%%%%%%%%%%%%%%%%%%%%%%%%%%%%%%%%%%%%%%%%%%%%%%%%%%%%%%%%%%%%%%%%%%%%%%%%%%%%%%%%%%%%
\subsection{Proof of convergence}\label{sec:proof1}
%%%%%%%%%%%%%%%%%%%%%%%%%%%%%%%%%%%%%%%%%%%%%%%%%%%%%%%%%%%%%%%%%%%%%%%%%%%%%%%%%%%%%%%%%%%%%%%%%%%%%%%%%%%%%%%%%

Before starting the discussion of convergence, we  shall briefly comment on an additional technical assumption made in   \cite{Caprini:2000js}. Specifically, in that paper it was assumed that the line rotated according to  (\ref{eq:steepline}) must not cross the real axis of the $u$ plane, in order to avoid hitting the singularities of the Borel transform. From this condition, the constraint $|\psi|<\pi/6$ was derived, where $\psi$ is the phase of $a$ (see Eqs. (36) and (39) of \cite{Caprini:2000js}). This is a rather strong constraint, but actually it turns out to be not necessary. Indeed, the integral  in (\ref{eq:In}) involves only the function $\tilde w(u)$, which has a branch point at $u=2$ and no other singularities for $u>0$. This means that, if the line of steepest descent  (\ref{eq:steepline}) reaches the axis $u=0$, it hits no singularities, but enters smoothly into the second Riemann sheet  of the function $\tilde w(u)$. Therefore, the constraint $|\psi|<\pi/6$ mentioned in  \cite{Caprini:2000js} is not necessary. 

We point out that extensive numerical calculations for mathematical toy models reported in \cite{Caprini:2009vf, Caprini:2011ya,  Caprini:2020lff,  Abbas:2013usa, Caprini:2019kwp} indicated convergence  in large regions of the coupling plane, not limited by the constraint $|\psi|<\pi/6$. The argument given above explains these results. Restrictions on the domain in the coupling plane arise only from the criteria of convergence discussed below.

 By inserting the expression (\ref{eq:Insaddle}) in (\ref{eq:cI}), $I(a)$ is written as a sum of two series, which in particular ensures the fact that the result is real when $a$ is real. For the study of convergence, we shall treat separately each of the two series. 
 Assuming,  as in  \cite{Caprini:2000js}, that a positive constant $c$ exists such that, at large $n$
  \be\label{eq:cn}
  |c_n|\approx e^{c\sqrt{n}},
  \ee
 we obtain the estimate
 \be\label{eq:estim}
 |c_n I^+_n(a)|\approx K n^{1/4} e^{- \xi \sqrt{n}}
 \ee
 where $K$ is a constant independent of $n$ and
  \be\label{eq:xi}
 \xi=\mbox{Re}[2^{3/4}(1+i) a^{-1/2}]-c.
 \ee
 
The convergence of the expansion (\ref{eq:cI}) has been studied in \cite{Caprini:2000js}
by considering the ratio
\be\label{eq:ratio}
\bigg\vert\frac{c_n I^+_n(a)}{c_{n-1} I^+_{n-1}(a)}\bigg\vert,
\ee
and requiring that it must be less than 1 for large $n$. However, it is easy to check that the limit of the ratio for $n\to\infty$ equals 1, and in this case the test is inconclusive,  the series may converge or diverge. Cauchy's root test is also inconclusive, since one can show that $\lim_{n\to\infty}|c_n I^+_n(a)|^{1/n}=1$.
 
The absolute convergence of the series can be established nevertheless using a direct comparison test. Namely, let us consider the inequality
\be\label{eq:comptest}
|c_n I^+_n(a)|\le \frac{1}{n^2},
\ee
which, using  (\ref{eq:estim}), is equivalent to
\be\label{eq:comptest1}
 K n^{9/4}\leq e^{\xi \sqrt{n}}.
 \ee
 This inequality  is clearly true for large $n$ if 
 \be\label{eq:condp}
 \xi=\mbox{Re}[2^{3/4}(1+i) a^{-1/2}]-c>0.
 \ee
 Since the series $\sum 1/n^2$ is absolutely convergent, the comparison test implies that the series
 $\sum c_n I_n^+(a)$ is also absolutely convergent, if the condition (\ref{eq:condp}) is satisfied.
 
 By treating in the same way the series  $\sum c_n I_n^-(a)$, we write finally the convergence condition in the compact form\footnote{This corrects two typos in Eq. (45) of \cite{Caprini:2000js}, where the  factor $2^{3/4}$ was missing and the sign in front of $c$ was wrong.}

 \be\label{eq:condpm}
\mbox{Re}[2^{3/4}(1\pm i) a^{-1/2}]-c>0.
\ee
 
 As noted in  \cite{Caprini:2000js},  if the coefficients $c_n$ grow less than any exponential,  $c_n<\exp[\epsilon \sqrt{n}]$ for an arbitrarily small $\epsilon$, the condition of convergence is 
 \be\label{eq:condpm0} 
\mbox{Re}[(1\pm i) a^{-1/2}]>0.
\ee
If, on the other hand, the coefficients grow faster than any $\exp[c \sqrt{n}]$, the series (\ref{eq:cI}) will be divergent. Note that such a behavior of $c_n$ is not excluded for expansions like (\ref{eq:Bw}), with radius of convergence equal to 1.

%%%%%%%%%%%%%%%%%%%%%%%%%%%%%%%%%%%%%%%%%%%%%%%%%%%%%%%%%%%%%%%%%%%%%%%%%%%%%%%%%%%%%%%%%%%%%%%%%%%%%%%%%%%%%%
\subsection{Generalizations}\label{sec:gen}
%%%%%%%%%%%%%%%%%%%%%%%%%%%%%%%%%%%%%%%%%%%%%%%%%%%%%%%%%%%%%%%%%%%%%%%%%%%%%%%%%%%%%%%%%%%%%%%%%%%%%%%%%%%%%%%%%
The arguments presented in the previous subsections can be easily generalized to other cases not treated in \cite{Caprini:2000js}. We consider first  the alternative expansion (\ref{eq:cItilde}), involving  a singularity-softening factor $S(u)$. Instead of
(\ref{eq:Inp}), we must evaluate now the  large-$n$ behavior of the quantity
 \be\label{eq:Inptilde}
\widetilde I^+_n(a)=\int\limits_{{\cal C}_+}  e^{-F_n(u)}\, \frac{d u}{S(u)},
\ee
where convenient choices for $S(u)$ are given in (\ref{eq:softu}) and (\ref{eq:softw}).

From the steps described in Sec. \ref{sec:steep}, it is clear that the main contribution to the integral is brought by the vicinity of the saddle point $u_0$. Since $S(u)$ is assumed to be a smooth function, we can apply the mean value theorem and factor out $1/S(u_0)$ in front of the integral. Then, instead of (\ref{eq:Inp2}), we have now
 \be\label{eq:Inp2S}
\widetilde I^+_n(a)\approx \frac{\zeta^n}{S(u_0)}  e^{-2^{3/4}(1+i) \sqrt{\frac{n}{a}}}
 \frac{e^{-i\beta/2}}{
 \sqrt{ |F_n''(u_0)|/ 2}}  \,\frac{\sqrt{\pi}}{ 2}.
\ee 
 From  (\ref{eq:softu}) and (\ref{eq:softw}) it follows that $S(u_0)$ behaves either as a power of $u_0$ or a constant. 
Recalling that $u_0\sim \sqrt{n}$ at large $n$, we can write,
up to a constant independent of $n$
\be\label{eq:InpsaddleS}
\widetilde I^+_n(a)\approx n^\delta \zeta^n
 e^{-2^{3/4} (1+i) \sqrt{\frac{n}{a}}},
\ee
 where $\delta$ is a real exponent. 
  One can use then this estimate in the direct comparison test, by simply adapting the arguments presented below (\ref{eq:comptest}).  Assume, like in (\ref{eq:cn}), that at large $n$ 
   \be\label{eq:cntilde}
  |\widetilde c_n|\approx e^{c\sqrt{n}},
  \ee
 where the coefficients $\widetilde c_n$ appear in (\ref{eq:BMiw}) and  (\ref{eq:cItilde}). It follows that the series (\ref{eq:cItilde}) converges in the domains described by (\ref{eq:condpm}) or (\ref{eq:condpm0}).
  
 In a similar way, one can establish the large-$n$ behavior of the quantities $\widehat I^+_n(a)$ defined in (\ref{eq:Inhat}), entering the FO expansion of the moments in the large-$\beta_0$ approximation. It is convenient to consider separately the two terms of $\sin\pi u=(e^{i\pi u}-e^{-i\pi u})/2i$, and combine them with the parameter $a$ by defining $1/\bar a =1/a \pm i\pi$. Then the steps presented in Sec. \ref{sec:steep},  performed with $a$ replaced by $\bar a$, lead to the estimate
\be\label{eq:InpsaddleM}
\widehat I^+_n(a)\approx n^\gamma \zeta^n
 e^{-2^{3/4} (1+i) \sqrt{n} \sqrt{{\frac{1}{a}}\pm i\pi}}.
\ee
Here the exponent $\gamma$ includes the contribution of the factor $P_i(u_0) S(u_0)$, which depends on the weight in the contour integral (\ref{eq:Mi})  and the softening factors, as seen in (\ref{eq:BMiw}). Using further the direct comparison test as in Sec. \ref{sec:proof1}, we can prove the convergence of the series (\ref{eq:cIhat}), provided the conditions
 \be\label{eq:condpmM} 
\mbox{Re}[2^{3/4}(1\pm i) (1/a\pm i\pi)^{1/2}]-c>0
\ee
 are satisfied, where $a=\beta_0 \alpha_s(s_0)/\pi$ and the constant $c$ is related by  (\ref{eq:cn}) to the behavior of the coefficients $c_n$,   or by (\ref{eq:cntilde}) to the behavior of the coefficients $\widetilde c_n$.
 
 Finally, it is easy to see that for the general FO expansion (\ref{eq:cIhati}), the condition of convergence will have the form (\ref{eq:condpm}), where $a=\beta_0 \alpha_s(s_0)/\pi$ and the constant $c$ is found from the growth of the coefficients $\hat c_{n,i}$ by
 \be\label{eq:hatcni}
 |\hat c_{n,i}|\leq \exp (c \sqrt{n}).
 \ee

%%%%%%%%%%%%%%%%%%%%%%%%%%%%%%%%%%%%%%%%%%%%%%%%%%%%%%%%%%%%%%%%%%%%%%%%%%%%%%%%%%%%%%%%%%%%%%%%%%%%%%%%%%%%%%%%%%%%%%%%%%%
\begin{figure}\vspace{0.8cm}
\includegraphics[width=5.5cm]{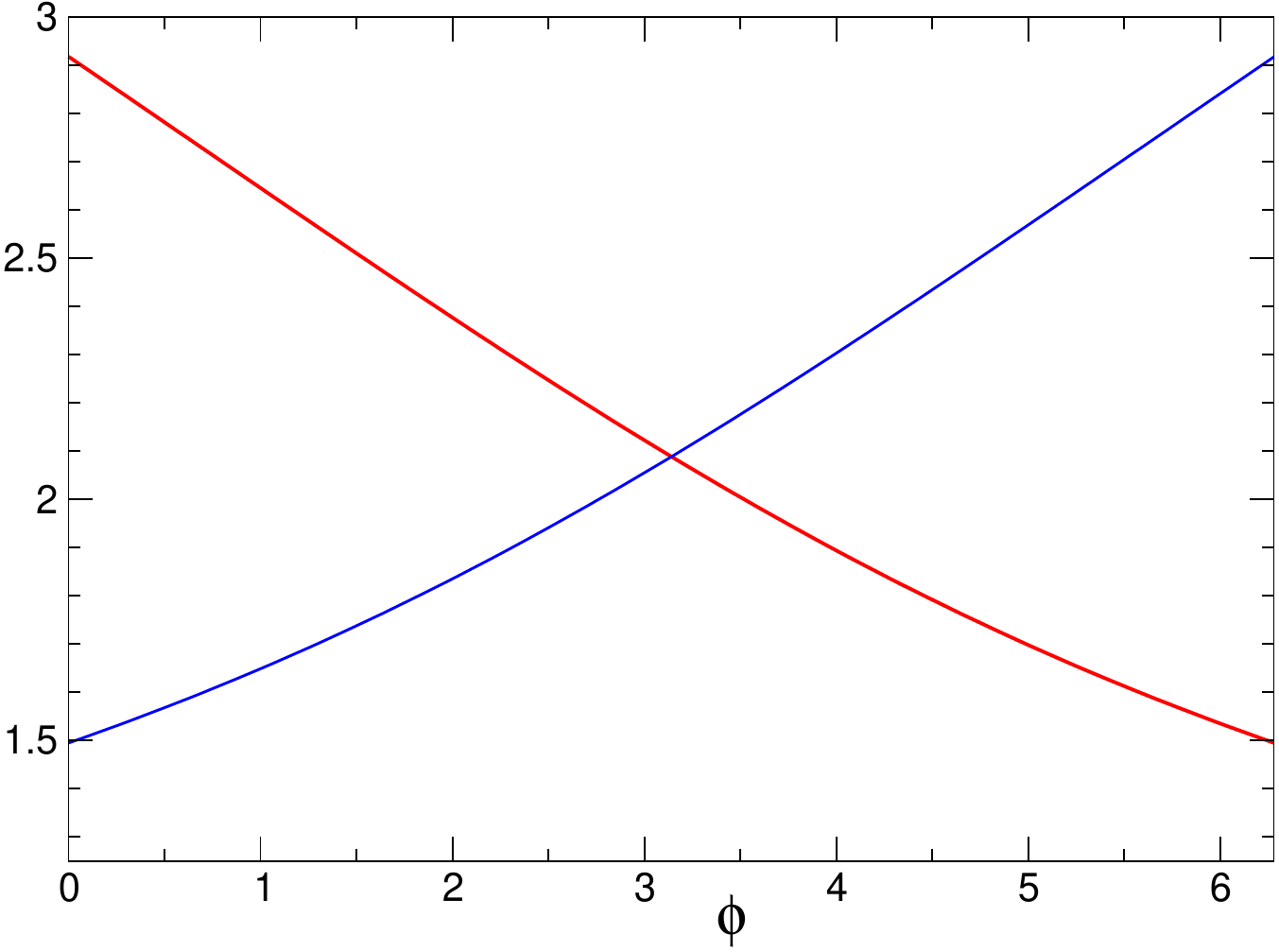}
\caption{The  quantities $\mbox{Re}[(1+ i) a^{-1/2}]$ (red) and $\mbox{Re}[(1- i) a^{-1/2}]$ (blue), for $a=\beta_0\alpha_s(-s)/\pi$ and  $s=m_\tau^2 e^{i\phi}$,  as functions of $\phi$. 
\label{fig:test0}}
\end{figure}
%%%%%%%%%%%%%%%%%%%%%%%%%%%%%%%%%%%%%%%%%%%%%%%%%%%%%%%%%%%%%%%%%%%%%%%%%%%%%%%%%%%%%%%%%%%%%%%%%%%%%%%%%%%%%%%%%%%%%%%%%%%%%%%%%%%%%%%%

%%%%%%%%%%%%%%%%%%%%%%%%%%%%%%%%%%%%%%%%%%%%%%%%%%%%%%%%%%%%%%%%%%%%%%%%%%%%%%%%%%%%%%%%%%%%%%%%%%%%%%%%%%%%%%%%%%%%%%%%%%%
\begin{figure}\vspace{0.8cm}
\includegraphics[width=5.5cm]{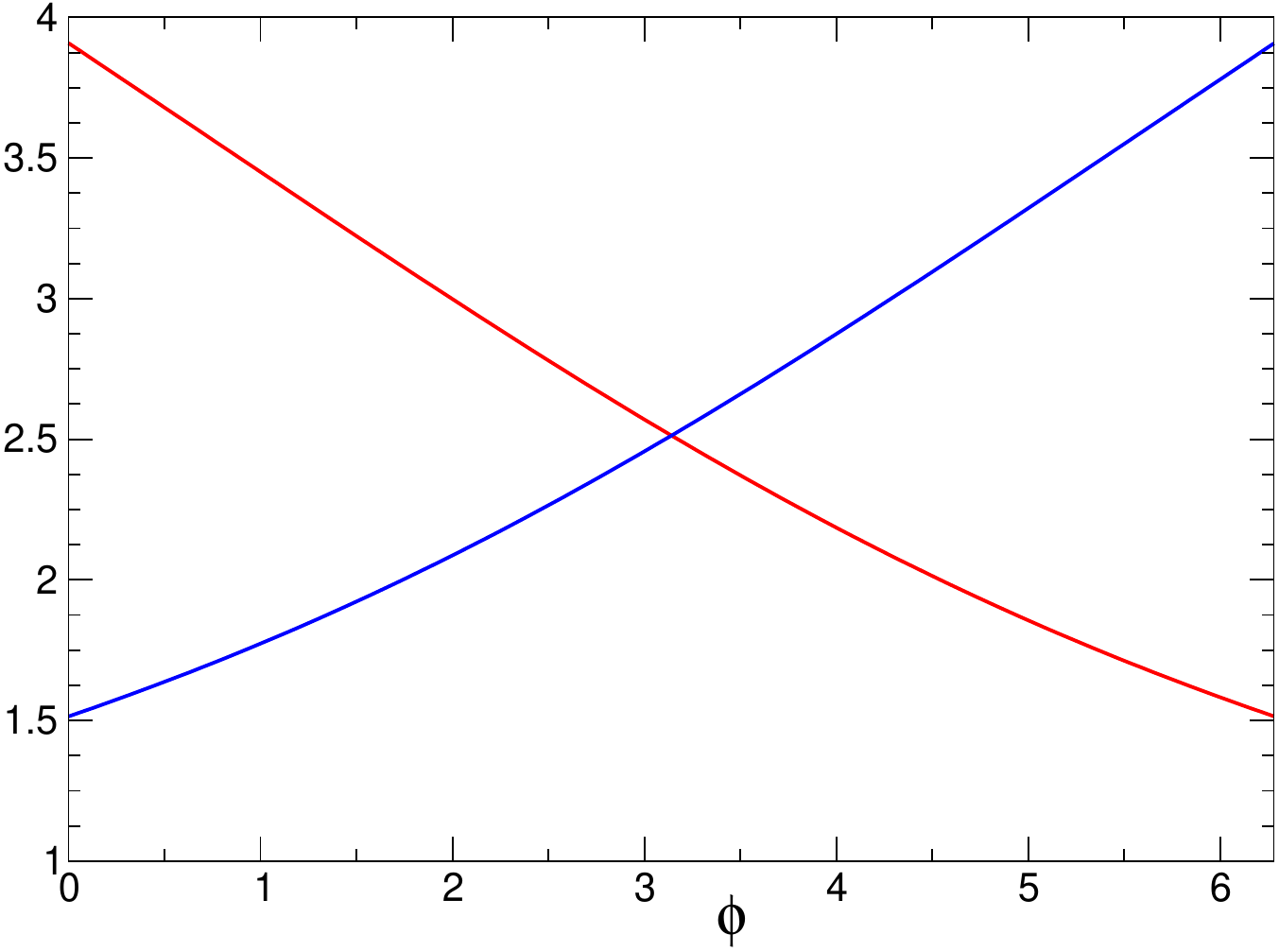}
\caption{The  quantities $\mbox{Re}[2^{3/4}(1+ i) a^{-1/2}]-1$ (red) and $\mbox{Re}[2^{3/4}(1- i) a^{-1/2}]-1$ (blue), for $a=\beta_0\alpha_s(-s)/\pi$ and  $s=m_\tau^2 e^{i\phi}$,  as functions of $\phi$. 
\label{fig:test1}}
\end{figure}

%%%%%%%%%%%%%%%%%%%%%%%%%%%%%%%%%%%%%%%%%%%%%%%%%%%%%%%%%%%%%%%%%%%%%%%%%%%%%%%%%%%%%%%%%%%%%%%%%%%%%%%%%%%%%%
\section{Convergence tests for the CI and FO expansions}\label{sec:applic}
%%%%%%%%%%%%%%%%%%%%%%%%%%%%%%%%%%%%%%%%%%%%%%%%%%%%%%%%%%%%%%%%%%%%%%%%%%%%%%%%%%%%%%%%%%%%%%%%%%%%%%%%%%%%%%%%%
In this section we shall investigate the fulfillment of the  convergence conditions established above for the expansions used in  the study of $\tau$ hadronic decay.
We consider first the perturbative expansion of the Adler function in the complex $s$ plane. As seen from  (\ref{eq:DI}), in this case the parameter $a$ is related to the running  coupling by $a=\beta_0 \alpha_s(-s)/\pi$. Therefore,  the conditions (\ref{eq:condpm}) or  (\ref{eq:condpm0}) can be viewed as defining regions of convergence of the perturbative expansion of the Adler function in the complex $s$ plane. For the calculation of the spectral moments,  it is of interest to check the validity of the convergence conditions  along the circle $|s|=m_\tau^2$.

As a first example, we take the Borel transform of a simple pole form $ B_{\widehat D}(u)=1/(2-u)$. In this case,  the coefficients  $c_n$ of the expansion (\ref{eq:Bw}) in powers of the conformal variable $w$ have the simple form $c_n=2n/3$, which grows less than any exponential at large $n$. Therefore, we must test the validity of the condition (\ref{eq:condpm0}). In the calculation,  we used the one-loop coupling from (\ref{eq:oneloop}), set $s_0=m_\tau^2$ and the value $\alpha_s(m_\tau^2)=0.32$, consistent with recent determinations (cf. \cite{PDG} and references therein).  In Fig. \ref{fig:test0} we plot the expressions in the l.h.s. of  (\ref{eq:condpm0}) calculated with this input along the circle  $|s|=m_\tau^2$. Both quantities are positive, as required by the convergence condition, which shows that the expansion (\ref{eq:cI}) of the Adler function is convergent along the circle  $|s|=m_\tau^2$, for a simple renormalon pole at $u=2$.

For a generic term of the form $1/(p-u)^\alpha$, with integer $p\ge 2$ and real $\alpha$, expected to be present in the  Borel transform $ B_{\widehat D}(u)$, the coefficients $c_n$ of the expansion (\ref{eq:Bw}) cannot be calculated  analytically exact in general. However, we checked numerically that they satisfy the condition $|c_n|< \exp(\sqrt{n})$ at large $n$. For instance, for $p=2$ and $\alpha=1.5$ the ratio $|c_n|/\exp(\sqrt{n})$ is equal to $5 \times 10^{-9}$ for $n=1000$ and to $4 \times 10^{-14}$ for $n=2000$. 
 For larger values of $p$, the growth of the coefficients $c_n$ slightly slows down. For instance, for $\alpha=1.5$ and $n=2000$, the  ratios $|c_n|/ \exp(\sqrt{n})$  are $4.5\times 10^{-19}$ for $p=3$,   $8\times 10^{-20}$ for $p=4$, and 
 $1\times 10^{-20}$ for $p=5$. 
 
We considered also negative values of $\alpha$, relevant for the expansion of the product $B_{\widehat D}(u) S(u)$   after softening  the lowest singularities as in  (\ref{eq:Bw1}). As expected,  because the residual singularity in the product is  mild, the growth of the coefficients $\widetilde c_n$ is less dramatic than for positive $\alpha$. For instance, for $p=2$ and $\alpha=-1.5$, the  ratios $|\widetilde c_n|/\exp(\sqrt{n})$  is equal to  $5\times 10^{-13}$ for $n=1000$ and to  $2\times 10^{-18}$ for $n=2000$.

From the numerical studies, we conclude that the conditions  $|c_n|< \exp(\sqrt{n})$ and  $|\widetilde c_n|< \exp(\sqrt{n})$  are satisfied at large $n$ for any finite sum of poles or branch points in the Borel transform $B_{\widehat D}(u)$.
The analysis presented in Sec. \ref{sec:proof}, shows that in this case convergence is ensured by the inequalities (\ref{eq:condpm}) with $c=1$.
 In Fig. \ref{fig:test0} we plot the expressions in the l.h.s. of  (\ref{eq:condpm}) for $c=1$, calculated with the one-loop coupling  along the circle  $|s|=m_\tau^2$. Both quantities are positive, as required by the convergence condition, which means that the expansions of the Adler function given in  (\ref{eq:DI})-(\ref{eq:Intilde}) are convergent along the circle, for the generic case of a Borel transform consisting from a finite sum of infrared renormalons. 
 
The above results imply that the CI expansions of the spectral moments are also convergent. Indeed, by inserting in (\ref{eq:Mi}) the relations (\ref{eq:DI})-(\ref{eq:Intilde})  and using the fact that the expansions (\ref{eq:cI}) and (\ref{eq:cItilde})   are absolutely convergent, we can permute the order of summation and integration and conclude that the CI expansion written in (\ref{eq:MicI}), and the similar one involving $\widetilde c_n$ and $\widetilde I_n$, are convergent. 

We note that the convergence of the CI expansions based on conformal mapping of the Borel plane for the Adler function in the complex $s$ plane and the moments was confirmed by numerical calculations on mathematical models in previous papers (see for instance Figs. 2, 4 and 8 from \cite{Caprini:2009vf}).

As concerns FOPT, we shall consider first the large-$\beta_0$ approximation, when  the expansions of the moments are defined by (\ref{eq:MiI})-(\ref{eq:Inhat}), with  coefficients $\widetilde c_n$ from the expansion (\ref{eq:Bw1}). As shown in Sec. \ref{sec:gen}, the convergence condition is represented by the inequalities (\ref{eq:condpmM}), where now $a=\beta_0 \alpha_s(m_\tau^2)/\pi$ and $c$ is the constant appearing in  (\ref{eq:cntilde}). From the above analysis of the Adler function, it follows that for Borel transforms with poles and branch points  we can take  $c=1$.  We checked  that  for this choice of $c$ and $\alpha_s(m_\tau^2)=0.32$  the inequalities  (\ref{eq:condpmM}) are satisfied  (the left sides are equal either to 3.9 or to 1.5). The conclusion of these tests is that the FO expansions  of the moments in the large-$\beta_0$ approximation, given in Eqs. (\ref{eq:MiI})-(\ref{eq:Inhat}),  converge for Borel transforms consisting from a finite sum of  infrared renormalons.
  
 We consider now the general FO expansion (\ref{eq:cIhati}), derived starting from (\ref{eq:MiFO}). These expansions include potentially large terms from the analytic continuation into the complex $s$ plane of
 the logarithms appearing in  (\ref{eq:hatD}), which may affect the convergence. This is confirmed by numerical calculations of the Adler function in the complex plane: see for instance Figs. 6 and 10 from \cite{Caprini:2009vf}, which show that the convergence is poor near the timelike axis. As a consequence, for the moments,  a good convergence is expected only if the weights  $\omega_i$ suppress this region.
 
  To check this expectation, we  considered as examples the kinematical weight $\omega_\tau(s/s_0)=(1-s/s_0)^3 (1+ s/s_0)$ and the weight  $\omega(s/s_0)=(1-s/s_0)^2$, which both vanish at $s=s_0$, and also the weight $\omega(s/s_0)=(1-2 s/s_0)$, which does not suppress the region near $s_0$.  The coefficients $c_{n,1}$ have been  generated by taking $B_{\wh D}(u)=1/(2-u)$.  The numerical calculations show that  the coefficients  $\hat c_{n,i}$ of the expansion (\ref{eq:Miw})  exhibit now a more pronounced increase,  and satisfy the inequality (\ref{eq:hatcni}) with $c=3$. For instance, the ratio $|\hat c_{35,i}|/\exp(3 \sqrt{35})$ is equal to 0.11 and 1.52 for the first two weights, and  to 15.6 for the third.  Unfortunately, for higher $n$ the accuracy of the calculations is no longer satisfactory, but the above values are an indication that for weigths suppressing the region near the timelike axis the convergence is better. Recalling that in this case the condition of convergence is the inequality  (\ref{eq:condpm}) with $a=\beta_0 \alpha_s(m_\tau^2)/\pi$, we checked that it is satisfied for $c=3$ (the l.h.s. is equal to 0.5). 
 
 Numerical calculations on mathematical models, performed in \cite{Caprini:2009vf, Caprini:2011ya,  Caprini:2020lff,  Abbas:2013usa},  confirm the tamed behavior of the FO expansions for moments with weights which suppress the region near the timelike axis.
  They confirm also that in this framework the CI expansions converge better. The reason is that the CI expansions implement  simultaneously the acceleration of the perturbative series and the renormalization-group improved coupling, while the FO expansions accelerate the perturbative series, but do not sum the  potentially large terms from the analytic continuation of logarithms into the complex plane. Thus, in the framework   based on the conformal mapping of the Borel plane, CIPT has a more solid theoretical basis. 
%%%%%%%%%%%%%%%%%%%%%%%%%%%%%%%%%%%%%%%%%%%%%%%%%%%%%%%%%%%%%%%%%%%%%%%%%%%%%%%%%%%%%%%%%%%%%%%%%%%%%%%%%%%%%%%%%%%%%%%%%%%%%%%%%%%%%%%%%%%%
\section{Summary and conclusions}\label{sec:conc} 
%%%%%%%%%%%%%%%%%%%%%%%%%%%%%%%%%%%%%%%%%%%%%%%%%%%%%%%%%%%%%%%%%%%%%%%%%%%%%%%%%%%%%%%%%%%%%%%%%%%%%%%%%%%%
In the present work we revisited the convergence of the modified QCD perturbative expansions based on the optimal conformal mapping of the Borel plane, proposed in \cite{Caprini:1998wg} and investigated further in  \cite{Caprini:2000js, Caprini:2001mn, Caprini:2009vf, Caprini:2011ya, Abbas:2013usa, Caprini:2019kwp, Caprini:2020lff, Caprini:2021wvf}. Our analysis brings some improvements to the proof of convergence presented in  \cite{Caprini:2000js}.  Thus, we showed that a technical assumption adopted in  \cite{Caprini:2000js} is not necessary, which leads to a considerably larger  domain of convergence. We also completed the proof given in  \cite{Caprini:2000js}, using instead of the ratio criterion, which gives inconclusive results at $n\to\infty$, the direct comparison test. Moreover, we generalized the proof to expansions with singularity softening, and to the perturbative expansions of the $\tau$ hadronic spectral moments. Finally, we performed a detailed analysis of the  convergence conditions (\ref{eq:condpm}), (\ref{eq:condpm0}) and  (\ref{eq:condpmM}), checking that they are satisfied along the circle $|s|=m_\tau^2$, for Borel transforms consisting from a finite number of poles and branch points. 
The results  are important because they provide a mathematical basis to the numerical calculations performed in previous papers \cite{Caprini:2009vf, Caprini:2011ya, Abbas:2013usa, Caprini:2020lff, Caprini:2021wvf}, where the behavior of the CI and FO expansions  was investigated up to orders of about 20 using models based on renormalons for generating the higher-order perturbative coefficients. 
 
The present work was motivated by the recent papers \cite{Hoang:2021nlz, Hoang:2020mkw,  Benitez-Rathgeb:2022yqb, Benitez-Rathgeb:2022hfj,  Gracia:2023qdy, Golterman:2023oml, Beneke:2023wkq}, which investigated the mathematical origin of the difference between  the FO and CI expansions of the $\tau$ hadronic  spectral moments, relating it to the sensitivity to the infrared renormalons\footnote{In \cite{Hoang:2021nlz, Hoang:2020mkw} it was even assumed that the CI and FO expansions correspond to different Borel sums, based on different prescriptions of regularizing the  ill-defined Laplace-Borel integral due to IR renormalons. We shall  not discuss here this assumption.}. In particular, in \cite{Benitez-Rathgeb:2022yqb, Benitez-Rathgeb:2022hfj,  Beneke:2023wkq},  the  discrepancy was solved by  subtracting the infrared renormalon divergence related to the gluon condensate, which amounts to a simultaneous  redefinition of the perturbative series and of the condensate. 

In this context, we thought to be useful to bring into attention the method of conformal mapping of the Borel plane, which amounts also to a redefinition of the perturbative series by exploiting the renormalons. Therefore, we can say that this approach is conceptually close to the methods proposed in  \cite{Benitez-Rathgeb:2022yqb, Benitez-Rathgeb:2022hfj,  Beneke:2023wkq}, although the practical implementation is different. We note in particular that the method of conformal mapping does not require the  normalization of the dominant infrared renormalon (the Stokes constant) and has no free parameters.

In the framework based on conformal mapping, the CI expansions and  the FO expansions (for moments with weights suppressing the region near the timelike axis)   exhibit a tamed asymptotic behavior, so the difference between their predictions is expected to be small, especially at high orders.  This feature was confirmed by previous numerical calculations on realistic models in \cite{Caprini:2009vf, Caprini:2011ya, Abbas:2013usa, Caprini:2020lff, Caprini:2021wvf}.  A similar behavior is obtained in \cite{Benitez-Rathgeb:2022yqb, Benitez-Rathgeb:2022hfj,  Beneke:2023wkq} after the subtraction of the gluon condensate renormalon (compare for instance  Fig. 3 from \cite{Caprini:2020lff} with Fig. 4 from \cite{Benitez-Rathgeb:2022yqb} and Fig. 2 from \cite{Beneke:2023wkq}). So, the method of conformal mapping of the Borel plane, as an alternative way of implementing information on renormalons in the perturbation series, is consistent with the methods proposed in \cite{Benitez-Rathgeb:2022yqb, Benitez-Rathgeb:2022hfj,  Beneke:2023wkq} for solving the CIPT-FOPT discrepancy.

We end with a few remarks about the nonperturbative corrections in the operator product expansion (OPE). In \cite{Hoang:2021nlz, Hoang:2020mkw,  Benitez-Rathgeb:2022yqb, Benitez-Rathgeb:2022hfj,  Gracia:2023qdy, Golterman:2023oml, Beneke:2023wkq} it was argued that  CIPT is  incompatible with the standard OPE. This is one of the reasons for which  FOPT was preferred already in  \cite{Beneke:2008ad, Beneke:2012vb}. The redefinition of the perturbative series proposed in  \cite{Benitez-Rathgeb:2022yqb, Benitez-Rathgeb:2022hfj, Beneke:2023wkq}, which solves the CIPT-FOPT discrepancy,  comes with a simultaneous redefinition of the OPE, in particular of the gluon condensate, such that both CI and FO expansions are consistent with OPE.
  
 In the approach based on conformal mapping, the original perturbative expansions in powers of the coupling are replaced by convergent series in terms of the expansion functions defined in  (\ref{eq:In}), (\ref{eq:Intilde}) and (\ref{eq:Inhat}) as Laplace-Borel integrals with  PV prescription. As we mentioned at the end of Sec. \ref{sec:conf}, these functions are singular at the origin of the coupling plane, exhibiting a nonperturbative behavior. Therefore, it is expected that the contribution of the additional nonperturbative terms, entering through the OPE, will be  different from those in the standard OPE. Actually, the fact that the method of conformal mapping  represents a realization of a renormalon-free OPE scheme, and in particular a renormalon-free gluon condensate scheme, was already  remarked in the literature (see footnote 8 of \cite{Benitez-Rathgeb:2022yqb}). The effective form of the OPE corrections to the perturbative expansions based on conformal mapping deserves further attention and will  be studied in a future work.

%%%%%%%%%%%%%%%%%%%%%%%%%%%%%%%%%%%%%%%%%%%%%%%%%%%%%%%%%%%%%%%%%%%

\end{document}